\documentclass[twocolumn,showpacs,preprintnumbers,amsmath,amssymb]{revtex4}

\usepackage{graphicx}
\usepackage{dcolumn}
\usepackage{bm}

\begin{document}

\preprint{APS/123-QED}

\title{Vortex formation of a Bose--Einstein condensate in a rotating deep optical lattice.}

\author{Akira Kato, Yuki Nakano, Kenichi Kasamatsu, and Tetsuo Matsui}
\affiliation{%
Department of Physics, Kinki University, Higashi-Osaka, Osaka 577-8502, Japan
}%
\date{\today}

\begin{abstract}
We study the dynamics of vortex nucleation and lattice formation in a 
Bose--Einstein condensate in a rotating square optical lattice 
by numerical simulations of the Gross--Pitaevskii equation. 
Different dynamical regimes of vortex nucleation are found, depending on the 
depth and period of the optical lattice.
We make an extensive comparison with the experiments by 
Williams {\it et al.} [Phys. Rev. Lett. {\bf 104}, 050404 (2010)], especially focusing on 
the issues of the critical rotation frequency for the first vortex nucleation 
and the vortex number as a function of rotation frequency. 
\end{abstract}

\maketitle
\section{Introduction}
Ultracold neutral atoms in an optical lattice (OL) are a particularly important system for 
studying a wide range of fundamental problems in condensed matter physics \cite{Bloch}. 
When an OL is rotated, the system mimics a lattice system of charged particles 
subject to a uniform magnetic field. This allows the development of versatile quantum simulators 
that can demonstrate various effects caused by a magnetic field 
such as quantum Hall effects \cite{Cooper,Dalibard}. 
Recently, two experiments have been reported making use of a rotating OL 
to study quantized vortex dynamics in gaseous Bose--Einstein condensates 
(BECs) \cite{Tung,Williams}. In this system, the vortex 
pinning parameters can be controlled by changing conditions such as  
the amplitude, lattice constant, and rotation frequency of the OL. 
Moreover, the realization of a synthetic magnetic field \cite{Lin}, simulated by 
the Raman process between internal states of atoms, opens the possibility of studying 
a wide range of phenomena caused by artificial magnetic fields 
well under control. 

Rotating BECs combined with a co-rotating OL exhibit a rich variety of 
vortex phases \cite{Reijnders,Pu,Sato,Kasamatsu2,Goldbaum,Mink}, 
which have two competing length scales: 
the two lattice spacings, one for the OL and the other for the vortex lattice. 
The central role of the OL is to pin vortices at its maxima. 
Tung {\it et al.} \cite{Tung} created a rotating square OL using a rotating mask, which
provided a periodic pinning potential that was stationary in the 
rotating frame associated with the presence of vortices. 
They observed a structural crossover from a 
triangular to a square lattice of vortices with increasing potential 
amplitude of the OL. 
In a deep OL, the condensates are well 
localized at each potential minimum, so that the system enters a regime  
which can be regarded as a Josephson-junction 
array \cite{Kasamatsu2,Trombettoni,Polini}. 
A bosonic Josephson-junction array under rotation 
(an analogue of a Josephson-junction array of superconductors 
under a magnetic field) can have 
characteristic vortex patterns with a unit-cell structure 
which depends on a filling factor, i.e., the number of vortices per unit cell of 
the periodic potential. 
This regime was recently demonstrated experimentally by Williams {\it et al.} \cite{Williams}. 

Motivated by the experiment of Williams {\it et al.}, we study the vortex lattice formation 
in a rotating BEC confined in a deep OL by numerical simulation of the 
Gross--Pitaevskii (GP) equation. We focus on the dynamics of 
vortices nucleation, which may depend on the properties of the OL, finding that 
the nucleation mechanism is greatly different from the surface instability in the case of a 
harmonically trapped rotating BEC \cite{Sinha,Tsubota,Parker}. 
An interesting experimental observation is that 
the minimum rotation frequency needed to nucleate a single vortex decreases 
as the amplitude of the OL is increased, and eventually falls below the rotation frequency 
at which a single vortex at the center of a condensate is energetically stable \cite{Williams}. 
This implies that vortices were nucleated locally at the lattice plaquette. 
However, our simulation reveals that this local nucleation does not occur 
for the experimental parameters used in Ref. \cite{Williams}. 
We also show that inclusion of a Gaussian envelope of the lattice beams is 
important to explain the experimental observation of the vortex number 
dependence on the rotation frequency.   
A preliminary numerical study on vortex nucleation in an OL was reported 
by Yasunaga and Tsubota \cite{Yasunaga}, 
but the case of a deep lattice limit was not discussed and sufficient understanding 
of the problem is still lacking. 

The paper is organized as follows. 
In Sec. \ref{formulation}, we describe the model of the problem 
and the parameter settings for our numerical simulations. 
We discuss the dynamics of the vortex lattice formation in a BEC 
under a rotating OL in Sec. \ref{dynamics} and make a detailed comparison 
with experiment in Sec. \ref{experiment}. 
Section \ref{conclusion} is devoted to a conclusion.

\section{Formulation of the problem} \label{formulation}
We study the dynamics of a BEC trapped in an external 
potential $V({\bf r})$ by employing the dissipative GP equation 
for the condensate wave function $\Psi({\bf r}, t)$ \cite{Tsubota,Choi,Penckwitt,Griffin}:
\begin{eqnarray}
(i - \gamma ) \hbar \frac{\partial \Psi}{\partial t} = \biggl[ -\frac{\hbar^2}{2m}\nabla^2 
+ V_{\rm ext} ({\bf r}) - \mu \nonumber \\
+ g |\Psi({\bf r},t)|^2 -\Omega L_{z} \biggr] \Psi({\bf r}, t). 
\label{GPequation}
\end{eqnarray}
Here, we take a frame rotating with frequency $\Omega$ around the $z$-axis. 
The total external potential $V_{\rm ext}$ is given by the sum of the 
harmonic potential $V_{\rm ho}$ and the OL $V_{\rm OL}$:
\begin{eqnarray}
V_{\rm ho} = \frac{1}{2} m \omega_{\perp}^{2} r^{2} + \frac{1}{2} m \omega_{z}^{2} z^{2}, \\
V_{\rm OL} =  V_0 \left[ \sin^2 (kx) + \sin^2 (ky) \right] \label{OLpote}
\end{eqnarray}
with $r^2 = x^2 + y^2$; we shall consider another form of $V_{\rm OL}$ 
later (see Eq. (\ref{OLpotegauss})).
. We have denoted that $L_z = - i \hbar (x \partial_{y} - y \partial_{x} )$ 
is the $z$-component of the angular momentum operator, 
$g = 4 \pi \hbar^2 a_s / m$ is the coupling constant with s-wave 
scattering length $a_s$, and $\gamma$ 
is the phenomenological dissipation parameter. The dissipation parameter is introduced 
to relax the system into the equilibrium configuration and is assumed to be 
$\gamma = 0.03$ in the following \cite{Tsubota,Choi}; we shall discuss the 
$\gamma$-dependence of the results later. 

To rewrite Eq.~(\ref{GPequation}) into a dimensionless form, 
we use the characteristic scales of the harmonic potential as 
length $a_{\rm ho} = \sqrt{\hbar / m \omega_{\perp}}$, time $\omega_{\perp}^{-1}$, 
and energy $\hbar \omega_{\perp}$. 
Since we are concerned with the vortex pattern and dynamics in the two-dimensional 
(2D) $x$--$y$ plane, we reduce Eq.~(\ref{GPequation}) into the 2D system. 
We assume that the profile of the wave function along the $z$-axis 
is approximately uniform by considering the system near the $z = 0$ plane.  
Then, the wave function can be decomposed as 
$\Psi (x,y,z,t)= \sqrt{N/R_{z}} \psi(x,y,t)$ with the normalization 
$\int dx dy |\psi|^2 = 1$, where $N$ is the total particle number and $R_z $
is a typical condensate size along the $z$-axis, taken as the Thomas--Fermi radius. 
The resulting dimensionless GP equation reads 
\begin{equation}
(i-\gamma) \frac{\partial \tilde{\psi}}{\partial \tilde{t}} = \biggl[-\frac{1}{2}(\partial^2_{\tilde{x}} 
+ \partial^2_{\tilde{y}}) + \frac{\tilde{r}^2}{2} + \tilde{V}_{\rm OL} - \tilde{\mu} 
+ \tilde{u}_{\rm 2D} |\tilde{\psi}|^2 - \tilde{\Omega} \tilde{L}_{z} \biggr] \tilde{\psi} 
\label{GPdimless}
\end{equation} 
with $\tilde{V}_{\rm OL} = \tilde{V}_{0} (\sin^2 \tilde{k} \tilde{x} 
+ \sin^2 \tilde{k} \tilde{y})$ and the dimensionless coupling constant 
$\tilde{u}_{\rm 2D} = 4 \pi N a_{s} / R_{z}$. Here, a dimensionless variable 
is denoted with a tilde, which shall be omitted in the following. 

We use the following parameter values to reproduce the experimental setup of Williams {\it et al.} \cite{Williams}: 
trapping frequencies 
$(\omega_{\perp}, \omega_{z})=2\pi (20.1, 53.0)$Hz, 
s-wave scattering length $a_{s} = 5.61$ nm, and 
particle number $N=1.0 \times 10^5$ yield $a_{\rm ho} = 2.4$ $\mu$m 
and $R_{z} = 5.6$ $\mu$m. 
Then, the coupling constant becomes $u_{\rm 2D} \approx 1000$. 
The spatial period $d=2$ $\mu$m of the OL used in the experiment 
corresponds to $d = \pi / k = 0.83 $ in our units. 
Hence, our free parameters are the potential depth $V_{0}$ in units of $\hbar \omega_{\perp}$
and the rotation frequency $\Omega$ in units of $\omega_{\perp}$. 
In the experiment \cite{Williams}, the range of $V_{0}$ is 
$100~\mathrm{Hz} \leq V_0 \leq 4000~\mathrm{Hz}$, 
corresponding to $5 \leq V_{0} \leq 200$ in our units. 
Nucleation of vortices was observed by starting from a non-rotating 
condensate loaded into both a harmonic trap and a static OL, 
and then turning on the rotation of the OL. We use the ground state solution of 
Eq.~(\ref{GPdimless}) in the presence of both the non-rotating harmonic trap and the OL 
as the initial state of the simulations. 
The numerical scheme to solve Eq.~(\ref{GPdimless}) is a Crank--Nicholson method 
with spatial mesh $\Delta_{x,y}=0.05$ and time step 
$\Delta_{t} = 0.0005$. Because the time development of Eq.~(\ref{GPdimless}) 
does not conserve the norm of the wave function for $\gamma \neq 0$, we 
treat the chemical potential $\mu$ as time-dependent and 
adjust it at each time step to ensure
normalization, by calculating the correction
$\Delta \mu = (\Delta t)^{-1} \ln [ \int d^2 r |\psi(t)|^{2} /\int d^2 r |\psi(t+\Delta t)|^{2} ]$ \cite{Jackson}. 


\section{Dynamics of a rotating BEC in an optical lattice}\label{dynamics}
The observation of vortex nucleation and lattice formation of a harmonically 
trapped BEC \cite{Madison2} ($V_{0} = 0$) has been well
reproduced by numerical simulations of the time-dependent 
GP equation \cite{Tsubota,Parker}. 
The simulation results clarified the interesting nonlinear dynamics 
where surface wave instability triggers vortex nucleation. 
We first show the overall nonlinear dynamics of vortex formation in a 
rotating BEC subject to an OL to reveal the mechanism of vortex nucleation. 
In this section, a rotation with $\Omega=0.6$ is suddenly started at $t=0$ of the simulations.   

\begin{figure*}
\centering
\includegraphics[width=1.02\linewidth]{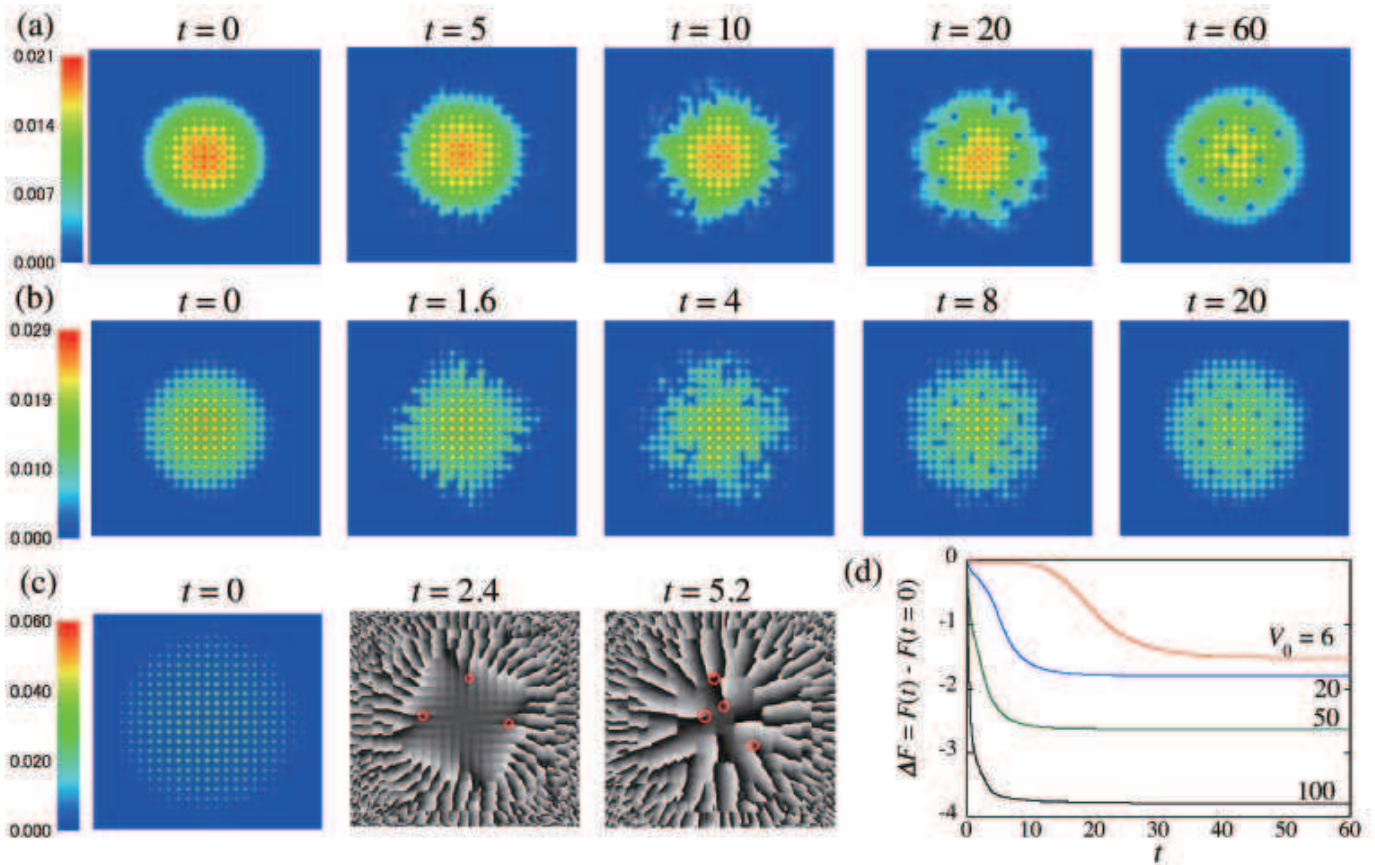}
\caption{(color online) Time development of the condensate density $|\psi|^2$ in the 
region $[-10,10] \times [-10,10]$ after the 
OL suddenly begins to rotate at $t=0$ with $\Omega=0.6$. The lattice constant of the 
OL is $d=0.83$ $(k = 1.2 \pi)$ and the height of the 
OL is $V_0 = 6$ (a), 20 (b), and 100 (c). In (c), the development of the phase 
$\theta = \arg \psi$ is also shown, where its value changes continuously from 0 (dark) 
to $2 \pi$ (bright). The vortices correspond to the ends of the branch cuts 
between the phases 0 and $2\pi$; some of them are marked by circles. 
In (d), time evolutions of the free energy for several values of $V_0$ are plotted 
as the difference from the free energy at $t=0$. The physical units are 
$a_{\rm ho} = \sqrt{\hbar / m \omega_{\perp}}$, $\omega_{\perp}^{-1}$, 
and $\hbar \omega_{\perp}$ for length, time, and energy, respectively.} 
\label{vortexdynamics}
\end{figure*}
Figure \ref{vortexdynamics} shows the time development of the condensate 
density for several values of the potential depth $V_{0}$. 
For the shallow OL $V_0 = 6$, the dynamics are similar to those without the OL, 
as shown in Fig.~\ref{vortexdynamics}(a). 
First, the surface of the condensate becomes unstable
and generates surface ripples that propagate along the surface.
Then, the surface ripples gradually develop into vortex cores, and these vortices
are pulled in toward the rotation axis to make a vortex lattice. 
The settled vortices are generally pinned by peaks of the OL.
For $V_{0}=20$ [Fig.~\ref{vortexdynamics}(b)], although vortices are also generated from 
the surface region, the condensate surface is disrupted to form density blobs 
rather than excited to form ripples. 
The vortices penetrate inside soon after the turn-on of the rotation. 
This is because quantized vortices accompany the density 
dips (vortex cores) so that they can easily penetrate into the condensate 
through the local density suppression caused by maxima of the OL. 
As the OL becomes much deeper, the system enters a regime 
of the Josephson-junction array. 
Here, fractions of the condensate are well localized at the potential minima 
and the overlap of the wave function between nearest-neighbor sites is very small. 
Then, the dynamics of the condensate density are completely frozen during the 
overall vortex nucleation process; only the phase is a dynamical degree of freedom. 
A typical example for $V_{0} = 100$ is shown in Fig.~\ref{vortexdynamics}(c). 
Even in this deep lattice, the vortices are nucleated from the periphery and settle 
very quickly as soon as the rotation is turned on. 

The different behavior in the characteristic vortex nucleation time can be seen 
from the development of the free energy $F=E-\mu-\Omega \langle L_z \rangle$, 
shown in Fig.~\ref{vortexdynamics}(d), where 
$E=\int d^2r \psi^{\ast} [-(\partial^2_x+\partial^2_y) +r^2/2+V_{\rm OL} +u_{\rm 2D} |\psi|^2] \psi$. 
For a shallow lattice, the energy stays constant for a while just after the rotation is turned on. 
During this period, the vortices near the surface are prevented from entering. 
After some time, the energy rapidly decreases through vortex 
penetration into the bulk. 
On the other hand, for a deep lattice, this period of constant energy cannot be seen. 
This means that the energy barrier for the vortex penetration is vanishingly small. 
Also, the vortices equilibrate soon due to the strong pinning effect of the OL. 
We checked that this qualitative feature of the decay process was not affected by 
the values of $\gamma$, which simply changed the decay time slightly for each numerical solution. 

\begin{figure}
\centering
\includegraphics[width=1.0\linewidth]{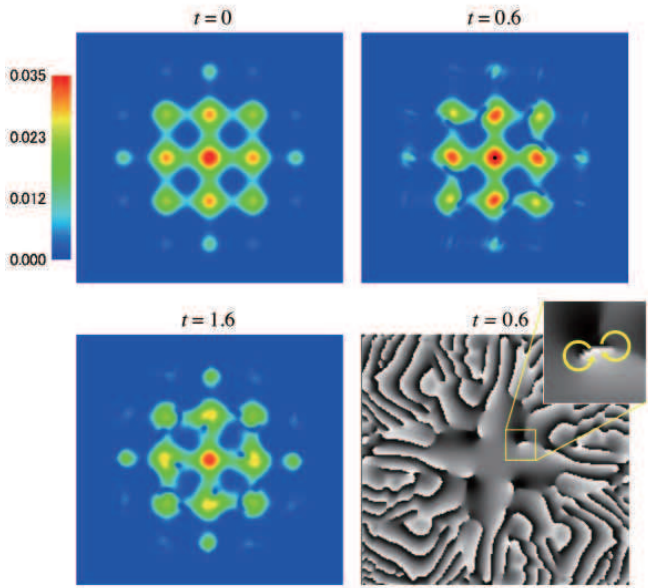}
\caption{(color online) Time development of the condensate density $|\psi|^2$ after the 
OL suddenly begins to rotate at $t=0$ with $\Omega=0.6\omega_{\perp}$ for $V_{0} = 20$ 
and $k=0.3\pi$ ($d =3.3a_{\rm ho}$) [compare to Fig.~\ref{vortexdynamics}(b)]. 
The bottom right panel shows 
the phase profile and the inset is an enlarged view of a vortex--antivortex pair.} 
\label{centernuc}
\end{figure}
Yasunaga and Tsubota \cite{Yasunaga} found a different kind of vortex nucleation dynamic, 
in which an OL potential generates vortex--antivortex pairs inside the bulk region. 
We find that this nucleation dynamic occurs for a relatively large lattice spacing $d$.  
For $V_{0} = 20$ but $k=0.3\pi$ ($d=10/ 3$), for example, a 
different dynamical feature from Fig.~\ref{vortexdynamics}(b) arises 
as shown in Fig.~\ref{centernuc}; some vortices are created from the surface as usual, 
while others arise from vortex--antivortex pairs created by peaks of the OL moving inside the
condensate. The phase profile at $t = 0.6$ reveals the pair creation occurring at
peaks of the OL [the bottom right panel of Fig.~\ref{centernuc}], 
because these vortices have circulations and
anti-circulations of the phase. Then, the anti-vortices quickly disappear by migrating 
outward and combining with
other vortices. However, some vortices remain in pairs in the condensate and
form a vortex lattice together with vortices coming from the condensate surface. 
We find that similar dynamics occur for different values of $V_0 \geq 20$.
This nucleation mechanism is similar to that of dragging the superflow 
through an obstacle potential \cite{Frisch} or, more closely, 
stirring the condensate by a circularly moving narrow potential \cite{Raman}. 

These different dynamical origins of vortex nucleation can be understood 
by noting that the density fluctuation is frozen for a deep OL. 
This condition can be obtained when the energy spacing 
$\sim (\pi\hbar/d) (2V_0/m)^{1/2}$ of a single well of the OL 
under the harmonic approximation becomes larger than the typical interaction energy $\sim \mu$. 
Then we have $d < d_{c} \equiv 2 \pi \xi \sqrt{V_0/\mu}$ for the density to freeze, where
$\xi = \hbar/\sqrt{2 m \mu}$ is the healing length. 
This estimation agrees fairly well with our 
numerical results because $d_{c} \sim 0.7$ in our parameters 
($\mu \approx 83$ and $\xi \approx 0.1$ for $V_0=100$ and $\Omega=0$). 
The lattice spacing $d = 2$ $\mu$m employed by Williams {\it et al}. corresponds to a regime 
in which the local vortex nucleation is not favorable because $d < d_c \sim 3 \mu$m 
for their parameters $V_{0}=2000$ Hz and $\mu=500$ Hz \cite{Williams}.

\section{Comparison with experimental observation}\label{experiment}
Here, we compare our results with the experiment by Williams {\it et al.} \cite{Williams}. 
Their significant observations are: 
(i) The minimum rotation frequency of the first vortex nucleation 
decreased with increasing $V_{0}$ and falls below $2 \pi \times 1$ Hz 
($\Omega=0.05$ in our physical units) 
above $V_0 \simeq 1500 $ Hz ($V_0 = 75$ in our physical units). 
(ii) For the deep lattice ($V_{0} > \mu$), the equilibrium vortex number 
increased linearly as a function of the rotation frequency $\Omega$. 
This point (ii) indicated that the condensate 
radius $R_{\perp}$ was not affected by the centrifugal expansion incidental 
to a rotating BEC in a harmonic potential, see Eq.~(\ref{feynmanrules}) below. 

\begin{figure}
\centering
\includegraphics[width=1.0\linewidth]{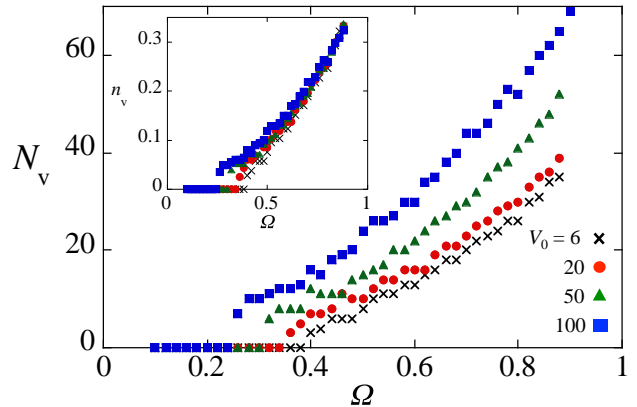}
\caption{(color online). 
Equilibrium vortex number as a function of $\Omega$ for several values of the potential depth $V_{0}$. 
We count the number of vortices within the radius determined by peripheral sites at which the density peak 
has 10\% of the central peak density. The inset shows the behavior of the vortex number density taken by the condensate radius 
at $\Omega=0$, i.e. $n_{\rm v} = N_{\rm v}/\pi R_{\perp}(\Omega=0)^{2}$.}
\label{vortexnumber}
\end{figure}
In Fig.~\ref{vortexnumber} we show the vortex number $N_{\rm v}$ in the equilibrium state 
as a function of the rotation frequency $\Omega$ for several values of $V_{0}$. 
This result was obtained by the dynamical simulation shown in the previous section 
with a sufficiently long time evolution to ensure the equilibration. We can see several features: 
(i) Naturally, the vortex number $N_{\rm v}$ increases monotonically with increasing $\Omega$. 
In contrast to the experiment, it grows faster than a linear function for all $V_{0}$.
(ii) In contrast with Ref. \cite{Williams}, the vortex number $N_{\rm v}$ also increases 
with increasing $V_{0}$ even at the same 
rotation frequency. This is because the tight confinement by the deep lattice effectively 
enhances the condensate radius $R_{\perp}$ under the condition of fixed 2D particle number \cite{tyuu}.
The vortex number thus increases with the radius because the lattice spacing of 
vortices is fixed by $\Omega$. However, this property is irrelevant to the $\Omega$-dependence 
of $N_{\rm v}$, because this 2D artifact only affects the radius at $\Omega=0$. 
As shown in the inset of Fig. \ref{vortexnumber}, when we plot the vortex density taken by 
the condensate radius at $\Omega=0$, i.e. $n_{\rm v} = N_{\rm v}/\pi R_{\perp}(\Omega=0)^{2}$, 
all the plots are almost coincident. This also supports the feature (i), namely, the scaling of $N_{\rm v}$ on 
$\Omega$ is independent of $V_0$.
(iii) The critical rotation frequency for the first vortex nucleation 
is decreased as $V_{0}$ increases.  
However, it is still above $\Omega=0.2$ even for the very deep lattice $V_{0} = 100$ 
in contrast with the experiment \cite{Williams}. In the following, we consider these 
issues in more detail. 
  
\subsection{Critical rotation frequency}
Our first concern is the critical rotation frequency of the vortex nucleation. 
To support our results, we calculate the thermodynamic critical frequency 
\begin{equation}
\Omega_c = \frac{E_1 - E_0}{\langle L_z \rangle} \label{thermocrieq}
\end{equation}
that ensures thermodynamic stability of the stationary state 
with a single vortex at the origin \cite{Fetterrev}. 
Here, $E_1$ and $E_0$ are the total energy for the single vortex state 
and the non-vortex state, respectively, and $\langle L_z \rangle$ is the mean 
angular momentum of the single vortex state. 
Since our OL of Eq.~(\ref{OLpote}) has a sine form which has a minimum at the origin, 
we replace it with a cosine form with a maximum at the origin, which is 
suitable for calculating the energy of the single vortex state pinned at the center. 
Note that this critical frequency is not relevant for the actual event of 
vortex nucleation seen in Fig.~\ref{vortexdynamics}, because the surface instability 
dominates the vortex nucleation. Nevertheless, we can obtain a lower 
bound of the critical rotation frequency because the critical rotation frequency $\Omega_{\rm sur}$ 
associated with the surface instability is typically larger than $\Omega_{c}$ of 
Eq.~(\ref{thermocrieq}) \cite{Sinha,Tsubota,Parker}. 

\begin{figure}
\centering
\includegraphics[width=1.0\linewidth]{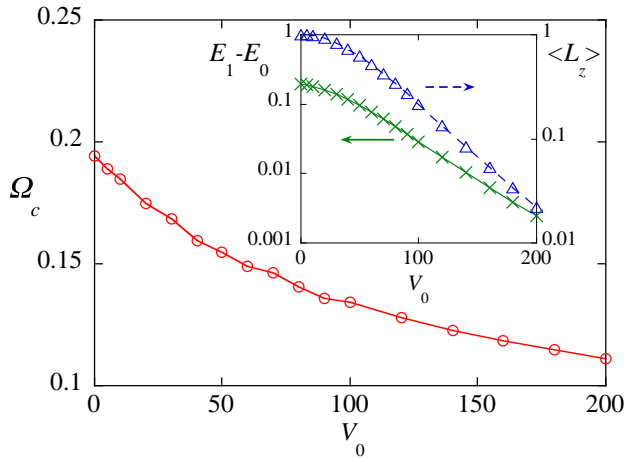}
\caption{(color online) Thermodynamic critical frequency $\Omega_c$ for 
the single vortex state as a function of the 
potential depth $V_0$. The inset shows the corresponding energy difference 
(numerator of the right-hand side of Eq. (\ref{thermocrieq})) and the mean 
angular momentum per atom (denominator of the right-hand side of Eq. (\ref{thermocrieq})). }
\label{crifre}
\end{figure}
Figure \ref{crifre} shows that the critical rotation frequency $\Omega_{c}$ 
decreases very slowly with $V_{0}$ and does not fall below 0.05 even at $V_0=200$. 
We find that, although the energy difference $E_{1} - E_{0}$ decreases exponentially 
with increasing $V_{0}$, 
$\langle L_{z} \rangle$ also decreases together so that their ratio $\Omega_{c}$ does not vanish. 
For sufficiently large $V_{0}$, there is little energy cost for a vortex to be positioned at the 
potential maximum, because an almost zero density region extends between the well separated 
condensate islands at the potential maxima. 
Correspondingly, the overlap of the wave functions between nearest neighbor sites decreases 
significantly, which leads to the suppression of the current flowing between the sites. 
This reduces the atoms contributing to the mean angular momentum. 

Thus, the dynamical simulations discussed in Sec.~\ref{dynamics} cannot account for 
the observation by Williams {\it et al.} \cite{Williams}, because
the vortices always nucleate from outside of the surface and are prevented 
from invading into the bulk for $\Omega < \Omega_c < \Omega_{\rm sur}$. 
Since no vortices were observed in a static OL ($\Omega=0$) \cite{private}, 
we can exclude the possibility of vortex nucleation due to thermal activation 
like the Berezenskii--Kosterlitz--Thouless mechanism \cite{Schweikhard} or 
interference between uncorrelated BECs \cite{Scherer}. 
Once vortices nucleate, they are expected to have a long life time 
because of the strong pinning effect of the OL, even if they are thermodynamically unstable. 
A more detailed study including the effect of the strong fluctuation of the OL \cite{Williams}, 
or applying a more elaborate model such as the projected GP equation \cite{Blakie} 
or stochastic GP equation \cite{Cockburn} to this problem is a challenge for future study.  

\subsection{Rotation frequency vs. vortex number}
The second concern is the linear dependence of the vortex number $N_{\rm v}$ 
as a function of the rotation frequency $\Omega$. 
For a superfluid rotating in a rigid container with radius $R$, 
the vortex number $N_{\rm v}$ can be determined by Feynman's relation 
$N_{\rm v} = m \Omega R^2 / \hbar$ \cite{Feynman}. 
In the Thomas--Fermi limit and if we smooth out the periodicity of the density caused 
by the OL, the radius of a harmonically trapped condensate under rotation is given by 
$R_{\perp}(\Omega)=R_{\perp}(\Omega = 0) [1 - (\Omega/\omega_\perp)^{2} ]^{-\nu}$ 
with $\nu=3/10$ for the 3D case and 1/4 for the 2D case \cite{Fetterrev}. 
Therefore, the radius expands as $\Omega \rightarrow \omega_{\perp}$ 
and the vortex number diverges as 
\begin{equation}
N_{\rm v} = \frac{m \Omega}{\hbar} R_{\perp}(\Omega)^2 
=  \frac{m \Omega}{\hbar} R_{\perp}(\Omega=0)^2 
\left[ 1 - \left( \frac{\Omega}{\omega_{\perp}} \right)^2 \right]^{-2 \nu}.
\label{feynmanrules}
\end{equation}
Contrary to this, Williams {\it et al.} observed that for a deeper lattice 
the vortex number increased linearly with increasing $\Omega$ and 
did not diverge at $\Omega=\omega_{\perp}$; 
the vortex number was nearly coincident with the prediction 
of Eq.~(\ref{feynmanrules}) without the factor 
$[1-(\Omega/\omega_{\perp})^2]^{-2 \nu}$, that is $N_{\rm v} \propto \Omega$. 
This result was expected to be caused by the effect of the OL, where
the radial expansion may be suppressed by the small tunneling rate 
at the peripheral lattice sites 
because the number of atoms decreases there. 

\begin{figure}
\centering
\includegraphics[width=1.0\linewidth]{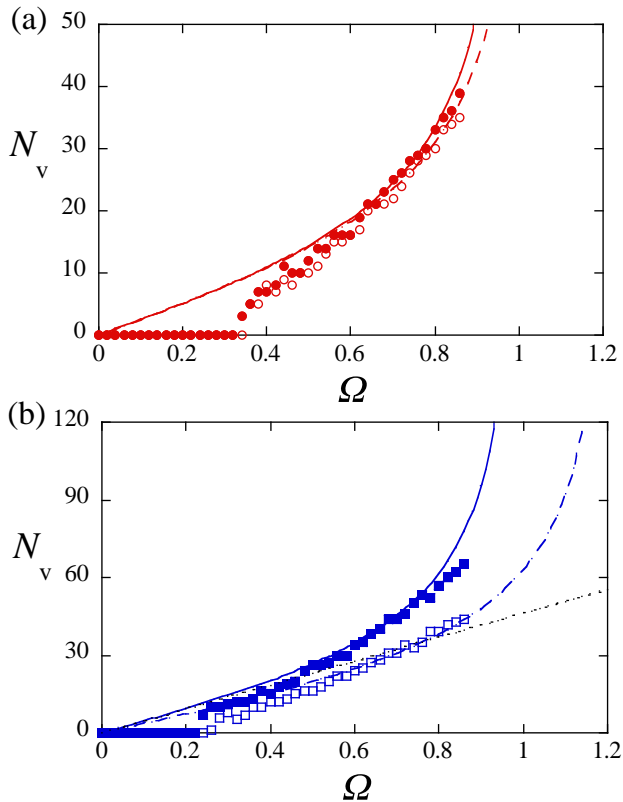}
\caption{(color online). Comparison of the vortex number with Feynman's rule 
for $V_0 = 20$ (a) and $V_0 = 100$ (b). 
The numerical results are plotted by filled symbols for the OL Eq. (\ref{OLpote}) and 
by empty symbols for the OL with a Gaussian envelope Eq. (\ref{OLpotegauss}). 
The solid and dashed curves correspond to 
Eq.~(\ref{feynmanrules}) and Eq.~(\ref{feynmanrules2}), respectively, with $\nu=1/4$. 
The fitting parameter $R_{\perp}(\Omega=0)$ is 5.0 for both curves in (a) and 
6.8 and 6.1 for the solid and dashed curves in (b). The dotted line in (b) 
represents $N_{\rm v}=R_{\perp}(\Omega=0)^{2} \Omega $ with $R_{\perp}(\Omega=0)=6.8$.
}
\label{vornumfit}
\end{figure}
We plot the vortex number taken from the numerical simulations as well as  Eq.~(\ref{feynmanrules}) 
in Fig.~\ref{vornumfit}. To this end, we regard the radius $R_{\perp}(\Omega=0)$ 
as a fitting parameter because it is a somewhat arbitrary value for a trapped BEC and 
our interest here is the scaling property of $N_{\rm v}$ with respect to $\Omega$. 
For $V_0=20$, where the OL plays a minor contribution, the vortex number can be well 
fitted by Eq.~(\ref{feynmanrules}), consistent with the experimental observation \cite{Williams}. 
A similar behavior can be seen for $V_0=100$ in Fig.~\ref{vornumfit}(b). 
Thus, we do not see a linear behavior of the vortex number as observed experimentally. 

In order to study the effect of an OL in more detail, we need a more realistic model of an OL. 
Here, we take into account the change in $V_{\rm OL}$ due to the 
Gaussian envelope of the lattice beams as 
\begin{eqnarray}
V_{\rm OL} =  - V_0 e^{-2 r^2/w^2} \left[ \cos^2 (kx) + \cos^2 (ky) \right] \label{OLpotegauss}
\end{eqnarray}
with beam waist $w=69$ $\mu$m \cite{AlAssam}. 
Since the Gaussian envelope behaves as $1-2r^2/w^2+{\cal O}(r^4/w^4)$, 
it leads to an increase in the frequency of the radial harmonic trap 
as $\sqrt{\omega_{\perp}^2+4V_{0}/mw^2}$. 
Under this modified potential, although the dynamical features of vortex nucleation are not 
altered, we can observe a significant reduction of $N_{\rm v}$ for large values of $V_{0}$, 
as shown by the empty symbols in Fig. \ref{vornumfit}. This reduction is caused by two contributions. 
The increase in the radial trap frequency makes the initial 
radius $R_{\perp}(\Omega=0)$ smaller than that without the Gaussian envelope. 
We found no visible change of $R_{\perp}(\Omega=0)$ for $V_{0}=20$, 
while $R_{\perp}(\Omega=0)$ becomes about 10\% smaller for $V_{0}=100$. 
Also, the Gaussian envelope suppresses the centrifugal expansion 
at $\Omega \to \omega_{\perp}$, which allows the condensate radius to remain 
finite even for $\Omega > \omega_{\perp}$, as observed in the experiment \cite{Williams}.
If we take up to the second order of $R_{\perp}/w$, Eq. (\ref{feynmanrules}) is modified as 
\begin{equation}
N_{\rm v} =  \frac{m \Omega}{\hbar} R_{\perp}(\Omega=0)^2 \left[ \frac{1 
- \left( \frac{\Omega}{\omega_{\perp}} \right)^2 
+ \frac{4V_{0}}{m \omega_{\perp}^2 w^2} }{1+\frac{4V_{0}}{m \omega_{\perp}^2 w^2}} \right]^{-2 \nu},
\label{feynmanrules2}
\end{equation}
which is plotted as the dashed curves in Fig. \ref{vornumfit}. 
Although only a minor modification occurs for small $V_{0}$, it is easy to see 
why the linear-like behavior was observed in the range $0<\Omega<1$ for large $V_{0}$. 
Our numerical results actually appear to follow the line of the linear scaling, 
as seen in Fig. \ref{vornumfit}(b). 

\section{conclusion}\label{conclusion}
We discuss the dynamics of vortex lattice formation in a BEC 
subject to a rotating OL. For a deep OL, we found that vortex nucleation 
occurs in a very different way from a usual rotating BEC in a harmonic trap.  
For the lattice spacing $d$ employed by Williams {\it et al.} \cite{Williams}, the 
motion of the density is frozen and vortices penetrate quickly from the outside 
through the density chinks. 
Local vortex nucleation, associated with the creation of vortex--antivortex pairs 
in the bulk region, occurs for lattice spacings larger than the critical value $d_c$. 
We also compare our numerical results with the experimental observations of Ref. \cite{Williams}, 
showing that the Gaussian envelope of the lattice beam is important in creating 
a difference in the scaling of the equilibrium vortex number on the rotation frequency.   

\begin{acknowledgments}
We thank R. A. Williams and S. Al-Assam for helpful comments. 
The work of K.K. is partly supported by a Grant-in-Aid for Scientific Research from JSPS (Grant No. 21740267).
\end{acknowledgments}

\begin{thebibliography}{99}
\bibitem{tyuu}
Intrinsically, $u_{\rm 2D}$ should be a nontrivial function of $V_0$ through the typical 
condensate size $R_z$ along $z$-axis, because the condensate can escape 
along the longitudinal ($z$-) direction with increasing $V_0$, which leads to 
increase in $R_z$. Thus, one can expect that $u_2D$ would be decreased with $V_0$.
\end{thebibliography}

\end{document}